\documentclass[]{achemso}
\usepackage[version=3]{mhchem} % Formula subscripts using \ce{}
\usepackage[T1]{fontenc}       % Use modern font encodings

\usepackage{amssymb}
\usepackage{amsmath}

\usepackage{color}

\usepackage[normalem]{ulem}
\usepackage{comment}

\author{Tianyu Su}
\affiliation[A]
{Department of Materials Science and Engineering, The Grainger College of Engineering, University of Illinois at Urbana-Champaign, 1304 W. Green Street, Urbana, Illinois 61801, United States}

\author{Brian J. Blankenau}
\affiliation[B]
{Department of Mechanical Science and Engineering, The Grainger College of Engineering, University of Illinois at Urbana-Champaign, 1206 W. Green Street, Urbana, Illinois 61801, United States}

\author{Namhoon Kim}
\affiliation[B]
{Department of Mechanical Science and Engineering, The Grainger College of Engineering, University of Illinois at Urbana-Champaign, 1206 W. Green Street, Urbana, Illinois 61801, United States}

\author{Jessica A. Krogstad}
\affiliation[A]
{Department of Materials Science and Engineering, The Grainger College of Engineering, University of Illinois at Urbana-Champaign, 1304 W. Green Street, Urbana, Illinois 61801, United States}

\author{Elif Ertekin}
\affiliation[B]
{Department of Mechanical Science and Engineering, The Grainger College of Engineering, University of Illinois at Urbana-Champaign, 1206 W. Green Street, Urbana, Illinois 61801, United States}
\alsoaffiliation[C]
{Materials Research Laboratory, University of Illinois at Urbana-Champaign, 104 South Goodwin Avenue, Urbana, Illinois 61801, United States}
\email{ertekin@illinois.edu}

\title{ Nitrogen-related short-range order in Fe-Ni-Cr austenitic stainless steels: first principles and cluster expansion study }
\begin{document}


\newpage
\begin{abstract}

Nitrogen (N) is a key alloying element that enhances the performance of Fe-Ni-Cr austenitic stainless steels, improving austenite stability, corrosion resistance, and yield strength. 
However, the role of N in modifying chemical ordering, particularly short-range order (SRO) and long-range order (LRO), is complex due to the multisublattice nature and magnetic interactions in these alloys.
In this work, we combine first-principles calculations with the spin cluster expansion (spin CE) method to systematically investigate the effects of N on chemical ordering in Fe-Ni-Cr alloys. 
Our atomistic models confirm a strong affinity between N and Cr, which drives the formation of N-Cr SRO and, at higher N concentrations, stabilizes $M_4$N-type ordered phases ($M$ = metal). 
Monte Carlo simulations reveal that low N concentrations promote local N-Cr or N-N SRO, while increasing N content leads to the emergence of Cr- and N-rich LRO structures. 
We also show that the presence of N suppresses intrinsic Fe-Cr and Ni-Cr SRO by competing with these interactions, particularly at high concentrations.  
The impact of Cr content on ordering diminishes as N approaches its solubility limit. These findings are consistent with experimental observations in high-N austenitic steels.
Finally, we discuss the influence of kinetic and magnetic effects on SRO evolution in high-N alloys. 
This study provides a comprehensive framework for understanding N-driven chemical ordering and offers insights into microstructural changes during nitriding processes.

\end{abstract}

\newpage 

\section{Introduction}

Nitrogen (N) is a widely used interstitial element that enhances the performance of many technologically important alloys, including high-entropy alloys \cite{he2023interstitial} and Fe-Ni-Cr-based austenitic stainless steels \cite{simmons1996overview,gavriljuk1999high}. 
In austenitic steels, N improves austenite stability, corrosion resistance \cite{masumura2015difference,baba2002role}, and mechanical strength via solid solution strengthening \cite{byrnes1987nitrogen}. 
Like carbon, N occupies octahedral interstitial sites in the face-centered cubic (FCC) lattice \cite{gavriljuk1999high} and tends to form interstitial-substitutional complexes that can significantly influence dislocation motion and mechanical behavior \cite{kawahara2024combined,jack1973invited}.
Experimental studies report planar dislocation slip in N-rich austenitic steels, suggesting that N alters both microstructure and deformation mechanisms \cite{mullner1993effect}. 
In particular, 
%\sout{enhanced slip planarity has been linked to the formation of short-range order (SRO) driven by nitrogen}
enhanced slip planarity is believed to be promoted by the nitrogen-driven short-range order (SRO) \cite{grujicic1992chemical,he2023interstitial}.
%\jak{Q: could this sentence be rephrased? as written is could sound like the localized slip causes SRO and not the other way around?--yes this is better, go ahead and make the change}  

Two main types of SRO have been identified in nitrogen-enriched austenitic stainless steels. 
The first involves interstitial–substitutional complexes in which N preferentially coordinates with Cr rather than Fe or Ni \cite{tong2019short,oddershede2010extended}. 
During nitriding, these interactions promote the formation of expanded austenite ($\gamma_N$), which forms without nitride precipitation. 
Strong Cr–N ordering in the $\gamma_N$ phase has been confirmed by extended X-ray absorption fine structure (EXAFS) analysis \cite{oddershede2010extended}, and nanoscale Cr–N clusters have also been observed by transmission electron microscopy \cite{xie2022nanosized}. 
However, as the nitrogen content approaches its solubility limit, $\gamma_N$ becomes thermodynamically unstable and may decompose into more stable phases, such as B1-type cubic CrN or martensite, under certain annealing conditions \cite{lei1999phase,li2003effect}.

The second form of SRO involves interactions between N atoms themselves, particularly the formation of second-nearest-neighbor (2NN) N–N pairs. 
Mössbauer spectroscopy studies of Fe–N alloys have revealed non-random distributions of N \cite{genin1973mossbauer}, with 2NN N–N pairs commonly observed and 1NN pairs rarely detected \cite{decristofaro1977interstitial,rochegude1986interstitial}. 
This distribution has been attributed to strong repulsion between 1NN N atoms, as supported by Monte Carlo simulations combined with Mössbauer data \cite{sozinov1999n}. 
The preference for 2NN configurations is consistent with the formation of ordered Fe$_4$N-type phases \cite{gavrilyuk1990distribution,sozinov1999n}. 
Such ordering is driven by strong electronic interactions between nitrogen and surrounding metal atoms, which increase free electron concentration and promote SRO stabilization \cite{gavriljuk2000correlation}.

From a mechanical standpoint, SRO plays a central role in shaping deformation behavior. 
Nitrogen-induced SRO promotes planar dislocation slip through a glide-plane softening mechanism \cite{gerold1989origin}. 
In particular, N–Cr SRO has been shown to suppress cross-slip and impede dislocation motion, %\sout{resulting in higher strain hardening rates in nitrogen-rich alloys}  
which can have a significant impact on the strain hardening rates in nitrogen-rich alloys \cite{gavriljuk1999high,kim2022effects,kawahara2024combined}. 
%\jak{I'd be a little careful with the preceeding two sentences.  For example, we typically think of suppressing cross-slip as a softening mechanims and impeding dislocation motion as a hardening mechanism.  I know that the literature is confusing here, but oversimplifying will only contribute to that confusion.  I'll try to think of some suggestions to rework this after I read all the way through.}
%\ts{From the Ref. 23 "dislocation cross-slip promotes dynamic recovery and reduces work hardening rate". So cross slip itself is a softening mechanism, right?}
%\jak{cross-slip can be a softening mechanism, but it is not always one.  cross-slip can also lead to greater dislocation entanglement when dislocation on different planes intersect and then this leads to hardening. I would rephrase it to say somethign like "...dislocation motion, which can have significant impact on the strain hardening rates in nitrogen-rich alloys." you're not dealing with dislocations in the body of this work, this way you can avoid a reviewer picking for something that's not really critical to your results.}
In contrast, Cr-rich nitrides contribute little to yield or tensile strength but are often associated with embrittlement \cite{jw1996effect}. 
These findings emphasize the importance of SRO evolution and microstructural control in tuning the mechanical performance of high-nitrogen alloys. 
A fundamental understanding of the origin and development of nitrogen-related SRO is thus essential for guiding alloy design and nitriding strategies.

Computational modeling offers valuable atomic-scale insight to complement experimental observations. 
Previous Monte Carlo simulations based on cluster variation methods have shown that nitrogen addition increases the population of Cr-rich octahedral lattice clusters \cite{grujicic1995models}. 
Recent studies further suggest that Cr atoms tend to arrange in pairs around N atoms, forming energetically favorable configurations \cite{tong2019short}. 
However, many of these investigations assume the presence of N–Cr SRO without systematically exploring how it evolves with changes in alloy composition, temperature, or magnetism. 
In practice, magnetic interactions among transition metal atoms significantly affect Cr ordering \cite{niu2015spin,walsh2021magnetically,su2024first}, which may in turn influence nitrogen distribution due to strong Cr–N affinity. 
Moreover, distinct magnetic states have been observed across nitrogen concentrations in high-N phases \cite{williamson1994metastable}, underscoring the complex interplay between magnetism and SRO evolution. 
The impact of nitriding on the intrinsic SRO among metal atoms also remains underexplored. 
A systematic, physics-based investigation of nitrogen-induced SRO under realistic thermodynamic and magnetic conditions is therefore essential for advancing the understanding and control of N-enriched austenitic steels.

To address these gaps, in this work we develop a spin cluster expansion (spin CE) framework that incorporates both chemical and magnetic degrees of freedom to study nitrogen-induced SRO in Fe–Ni–Cr austenitic stainless steels. 
Using Monte Carlo simulations, we investigate three types of SRO: metal–metal, N–N, and N–metal. 
Our results reproduce key experimental observations, including the formation of N–N 2NN pairs across a broad range of nitrogen concentrations and the preferential emergence of N–Cr SRO over N–Fe and N–Ni at elevated temperatures. 
At high nitrogen content, N incorporation strongly perturbs the intrinsic metal SRO, favoring the formation of M$_4$N-like structures—consistent with experimental findings. 
Additionally, we show that pre-existing metal SRO can influence nitrogen distribution when the metal sublattice is fixed, and that the magnetic state of the alloys can influence the formation of Cr nitride.
These results reveal the coupled effects of composition, temperature, and magnetism in governing SRO formation in high-nitrogen austenitic alloys.

\section{Methods}

\subsection{Spin cluster expansion}

Building on our previous work~\cite{kim2022multisublattice,su2024first}, we developed a multisublattice, multicomponent spin CE (CE) framework to model the thermodynamics of high-N austenitic stainless steels.
Conventional CE models~\cite{Sanchez1984,van2002alloy} use orthogonal basis functions to represent configuration-dependent properties, but in multicomponent systems, extracting physically meaningful interactions, especially for triplet and higher-order clusters, requires transformation of the effective cluster interactions (ECIs)~\cite{wolverton1994cluster,wrobel2015phase,kim2022multisublattice}.
Further, most prior CE models either neglect magnetism or treat it implicitly via magnetic ground-state fitting.

To overcome these limitations, we employ a modified spin CE formalism that decouples chemical and magnetic interactions.
The total energy is expressed as:
\begin{equation}
    E_{CE}(\vec{\sigma}) = \sum_{\alpha} J_{\alpha} \Theta_{\alpha} (\vec{\sigma}) + \sum_{\beta} \sum_{\langle i,j \rangle} J_{\beta} S_{i} S_{j} \hspace{1em}. \label{ce2}
\end{equation}
The first term describes chemical interactions across clusters $\alpha = {\alpha_1, ..., \alpha_l}$ of size $l$, where $J_{\alpha}$ is the ECI and $\Theta_{\alpha}(\vec{\sigma})$ counts its occurrences in configuration $\vec{\sigma}$. 
The second term captures magnetic exchange interactions between spin pairs $\langle i,j \rangle$, with $J_{\beta}$ denoting the exchange constant for dimer $\beta$ and $S_i$, $S_j$ representing spin values.

A large number of candidate clusters, including metal–metal, metal–nitrogen, and nitrogen–nitrogen, were generated based on a cutoff radius. 
To identify the most relevant terms, we employed a compressive sensing approach~\cite{nelson2013compressive} and solved the following LASSO-regularized regression: 
\begin{equation}
    J = arg \min_{J} \left\{ \frac{1}{N} \sum_{i=1}^N ( E_{i, DFT} - E_{i, CE}(\vec{\sigma}) )^2 + \lambda \sum \left| J \right| \right\} \hspace{1em}.
	\label{lasso}
\end{equation}
Before fitting, all features were normalized to zero mean and unit variance to ensure numerical stability. 
Model selection and regularization were guided by 10-fold cross-validation, and the parameter $\lambda$ was tuned to balance sparsity and predictive accuracy.
Final CE models were validated by benchmarking against experimental observations, as discussed in the Results section.

\subsection{Monte Carlo Simulations}

We implemented a lattice Monte Carlo method in the canonical ensemble based on the spin CE framework. 
To maintain constant composition throughout the simulations, Kawasaki dynamics were employed to facilitate atomic swaps between lattice sites~\cite{kawasaki1966diffusion}. 
Each MC step consisted of either an atom swap or a spin-flip trial, chosen with equal probability, enabling simultaneous equilibration of both chemical and magnetic degrees of freedom.

Simulations were carried out over a temperature range of 500 K to 1500 K, in 100 K increments. 
To ensure proper mixing, initial disordered configurations were equilibrated at the highest temperature (1500 K), followed by sequential cooling and equilibration at each lower temperature. 
The Metropolis-Hastings algorithm~\cite{metropolis_1953} was used to sample configurations from the Boltzmann distribution.

A $10 \times 10 \times 10$ conventional FCC supercell, containing 4000 metal atoms with varying nitrogen contents, was used to mitigate finite-size effects and ensure energy convergence within 0.1 meV/atom. 
At each temperature, we performed 2000 MC passes per atom for equilibration, followed by 6000 passes to evaluate thermodynamic properties. 
Convergence tests confirmed that this sampling scheme was sufficient to reach equilibrium within 0.1 meV/atom.

The degree of chemical SRO was quantified using the Warren–Cowley parameter:
\begin{equation}
    \alpha_l^{AB} = 1 - \frac{P_l^{AB}}{C_AC_B} = 1 - \frac{p_{l,A}^B}{C_B} \hspace{0.5em}, 
\label{WC_SRO}
\end{equation} 
where $P_l^{AB}$ is the probability of finding an $AB$ pair in the $l$-th neighbor shell, and $p_{l,A}^B = P_l^{AB} / C_A$ is the conditional probability of finding species $B$ in the $l$-th shell surrounding an atom of species $A$. 
Here, $C_A$ and $C_B$ denote the concentrations of species $A$ and $B$, respectively. 
For a completely random solution, $\alpha_l^{AB} = 0$. 
A positive value of $\alpha$ indicates a preference for like-pair clustering ($AA$ or $BB$), while a negative value suggests a tendency toward unlike-pair ordering ($AB$). 

Although originally formulated for metallic alloys, the Warren–Cowley SRO formalism can be extended to nitrogen-containing alloys. 
The N–metal SRO parameter is defined as:
\begin{equation}
    \alpha_l^{N-M} = 1 - \frac{n_{l}^M}{K_{l}C_M} \hspace{0.5em}, 
\label{N_metal_SRO}
\end{equation} 
where $n_l^M$ is the number of metal atoms $M$ in the $l$-th neighbor shell of a nitrogen atom, $K_l$ is the coordination number of that shell, and $C_M$ is the concentration of element $M$. 
The ratio $n_l^M / (K_l C_M)$ represents the conditional probability of finding metal atom $M$ around a nitrogen atom.

Similarly, the N–N SRO parameter is defined as: 
\begin{equation}
    \alpha_l^{N-N} = 1 - \frac{n_{l}^N}{K_{l}C_N} \hspace{0.5em}, 
\label{N_N_SRO}
\end{equation} 
where $n_l^N$ is the number of nitrogen atoms in the $l$-th neighbor shell of a given nitrogen atom, and $C_N$ is the nitrogen concentration, defined as the nitrogen-to-metal (N/M) ratio. 
Because nitrogen and metal atoms occupy different sublattices, the 1NN N–N and 1NN N–M pairs reside in different neighbor shells (see Figure~\ref{fig1}(a)). 

The values of N-related SRO parameters can range from negative values to 
1. In cases where a particular N–M pair is highly favored, the conditional probability can approach unity. 
If the concentration of metal species $M$ is low, this can yield a highly negative SRO parameter, potentially well below –1.

\subsection{First-principles data generation}

Spin-polarized density functional theory (DFT) calculations were performed using the Projector Augmented-Wave (PAW) method~\cite{kresse1999ultrasoft}, as implemented in the Vienna Ab initio Simulation Package \cite{kresse1993ab,kresse1996efficient}. 
The exchange-correlation interactions were approximated using the Perdew-Burke-Ernzerhof (PBE) functional\cite{perdew1996generalized}, and PAW-PBE pseudopotentials with frozen semicore states were employed. The valence configurations were set as [Ar]3d$^7$4s$^1$ for Fe, [Ar]3d$^9$4s$^1$ for Ni, [Ar]3d$^5$4s$^1$ for Cr, and [He]2s$^2$2p$^3$ for N. 

The plane-wave energy cutoff was set to 500 eV. 
Fermi-level smearing was applied using the first-order Methfessel–Paxton method with a width of 0.05 eV. 
A $k$-point density of 2400 points per reciprocal atom was adopted, corresponding to an $11 \times 11 \times 11$ Monkhorst–Pack grid for the primitive FCC cell, and scaled appropriately for larger supercells. 
Convergence tests confirmed that this sampling scheme ensured energy convergence within 0.4 meV/atom. Structural relaxations were performed until the total energy converged to within $10^{-6}$ eV/cell and forces were reduced below 0.02 eV/Å.  
Multiple initial magnetic configurations were assigned to each structure to capture magnetic degrees of freedom, following the approach in our previous work~\cite{su2024first}.

We focused on FCC Fe–Ni–Cr alloys relevant to austenitic stainless steels. 
Initial structures without nitrogen were generated using the Alloy Theoretic Automated Toolkit (ATAT) \cite{van2002alloy} via a variance-reduction approach tailored for the Fe–Ni–Cr ternary system. 
To introduce nitrogen, we employed $2 \times 2 \times 2$ special quasi-random structures (SQS) with varied Fe–Ni–Cr compositions, generated using the \texttt{mcsqs} code\cite{van2013efficient}. 
Nitrogen atoms (1, 2, 4, or 8 total) were placed in interstitial sites of these SQS to span a range of concentrations. 
The corresponding alloy compositions are listed in Figure S1. 

During CE model development, several ground states and high-N phases were identified, including the cubic B1 CrN phase. 
Additional nitrides and high-N intermetallic compounds, such as Fe$_2$NiCrN, Fe$_3$CrN, FeNi$_3$N, FeN, CrN, and NiN, were incorporated into the training dataset to enhance model accuracy and predictive capability.

\section{Results}

\subsection{Construction and Benchmark of the spin CE Model}

\begin{figure*}[!hbtp] %*
\centering
\includegraphics[width=3in]{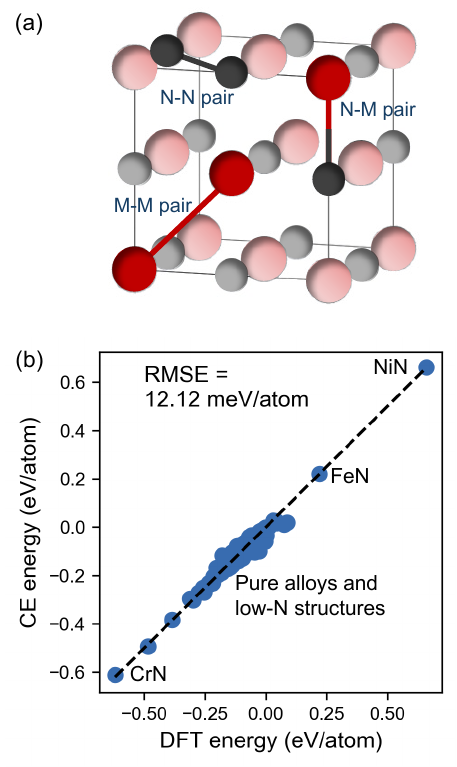}
\caption{ 
(a) Schematic illustration of the two face-centered cubic (FCC) sublattices in the Fe–Ni–Cr–N system. Red spheres represent atoms on the metal sublattice (Fe, Ni, Cr), while black spheres represent nitrogen atoms on the interstitial sublattice. Only pairwise interactions are shown for clarity, though the model also includes many-body (trimer and quadrumer) interactions.
(b) Comparison between DFT-calculated formation energies and predictions from the cluster expansion (CE) model for a set of Fe–Ni–Cr–N structures. The two high-energy structures correspond to NiN and FeN nitrides, while the lowest-energy structure is CrN nitride.
}
\label{fig1}
\end{figure*}

We begin by constructing a cluster expansion (CE) model for the FCC Fe–Ni–Cr–N system. 
Building on our previous approach~\cite{su2024first}, we include spin clusters among metal atoms up to third-nearest-neighbor (3NN) distance and optimize the set of chemical clusters to balance model complexity and predictive performance. 
The chemical interactions in the model include metal–metal, metal–nitrogen, and nitrogen–nitrogen interactions, as illustrated in Figure~\ref{fig1}(a).

While conventional performance metrics such as root mean square error (RMSE) and cross-validation (CV) error—are useful for evaluating model accuracy, our earlier work~\cite{su2024first} showed that models with similar RMSE values can still differ significantly in their predictive capabilities. 
Therefore, benchmarking against experimental and theoretical data, such as SRO parameters, order–disorder transition temperatures, and Curie temperatures, is essential for validating the model’s reliability. 
A comprehensive benchmark dataset established in Ref.~\cite{su2024first} is used to assess performance, and the final model is selected based on its agreement with the majority of these benchmarks.

The final spin CE model includes 15 pair clusters (dimers), 34 three-body clusters (trimers), and 3 four-body clusters (quadrumers). 
Detailed information on cluster geometries and their effective cluster interactions (ECIs) is provided in Table S1 and Figure S2 of the Supplementary Information (SI). 
Notably, the dominant features of the ECIs include strong negative interactions for Cr–N 1NN pairs and positive interactions for N–N 1NN pairs, features that are consistent with known experimental observations and thus lend further confidence to the model.

Figure~\ref{fig1}(b) compares the DFT-computed formation energies with CE predictions, yielding an overall RMSE of 12.12 meV/atom. 
While the model successfully captures key energetic trends across the alloy system, it initially failed to reproduce the Curie temperature of elemental Ni. 
To address this, we applied constraints to the 2NN and 3NN Ni–Ni spin clusters, yielding a refined model that accurately captures Ni’s magnetic behavior. 
Importantly, these adjustments had minimal impact on the predicted SRO behavior (see SI Figures S3–S5 for full benchmark results). 
As a result, all SRO predictions reported in this work are based on the original, unconstrained CE model, except for the analysis of magnetic effects on SRO as noted later.

 %*

\subsection{N-related SRO at Different N Concentrations}

We employed the optimized cluster expansion (CE) model to calculate Warren–Cowley SRO parameters for N–metal and N–N pairs in Fe–Ni–Cr–N alloys. 
To represent a typical high-nitrogen austenitic stainless steel, we selected an alloy composition of Fe$_{70}$Ni$_{10}$Cr$_{20}$, incorporating either 1 at.\% or 10 at.\% N atoms (corresponding to the N/M ratio).
Note that the 10 at.\% (approximately 3 wt.\%) N concentration is the solubility limit of N in Fe-Ni-Cr alloys at ambient conditions~\cite{frisk1991thermodynamic,frisk1992solubility}.
However, after surface nitriding, the near-surface nitrogen content can substantially exceed this limit, reaching values around 40 at.\%~\cite{somers2013nitriding}.
%\jak{Does this composition range need to be addressed? 1 at\% is very high for most conventional compositions, and at such high levels it is almost always in some form of precipitate.}
%\ts{Yes, maybe I should say that we want to study high-N alloys. 1 at\% is around 0.25 wt\%, which is pretty close to the KU alloys (0.3 wt\%). 10 at\% is the solubility limit.}
%\jak{great!  i think that context is important.}

\begin{figure*}[!hbtp] %*
\centering
\includegraphics[width=6in]{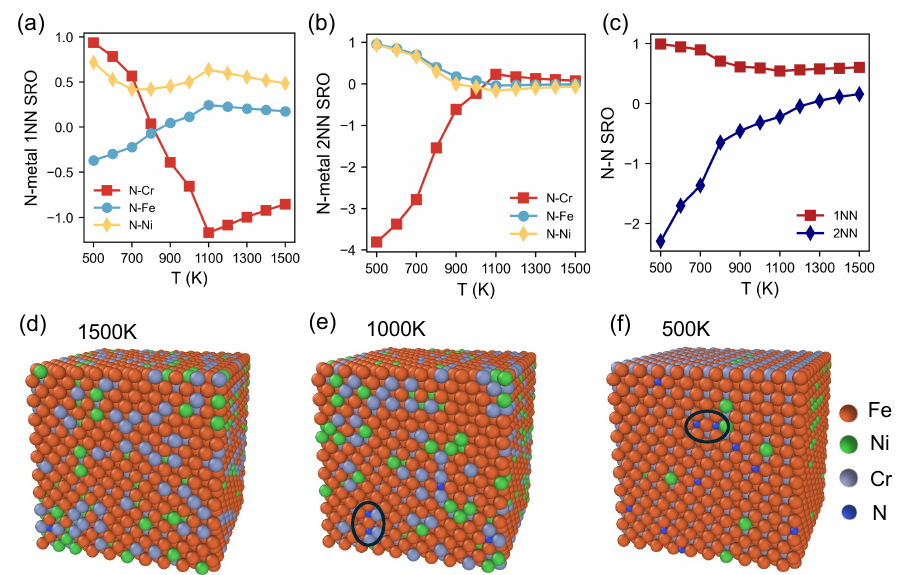}
\caption{
(a, b) First- and second-nearest-neighbor (1NN and 2NN) SRO parameters between nitrogen and metal atoms, and (c) 2NN N–N SRO parameter as a function of temperature for the 1 at.\% N case in Fe$_{70}$Ni$_{10}$Cr$_{20}$.
(d–f) Representative Monte Carlo snapshots at 1500 K, 1000 K, and 500 K, respectively, showing the atomic distributions. Black circles in (e) and (f) highlight regions exhibiting N–N 2NN short-range order.
Fe, Ni, Cr, and N atoms are shown in red, green, blue, and dark blue, respectively.
}
\label{fig2} 
\end{figure*}

For the 1 at.\% N case, the temperature-dependent SRO parameters for N–metal and N–N pairs are shown in Figure~\ref{fig2}. 
As illustrated in Figure~\ref{fig2}(a), the 1NN N–Cr SRO parameter exhibits a marked transition from negative to positive around 800 K. 
This trend suggests that N–Cr 1NN pairs are increasingly favored at high temperatures, whereas N–Fe 1NN pairs dominate at lower temperatures. 
Interestingly, Figure~\ref{fig2}(b) shows that the N–Cr 2NN SRO parameter remains negative below 1000 K, indicative of order–disorder transition behavior. 
% \jak{Is there a typo in the preceding sentence? Figure 2b shows the 2NN SRO parameter, right? It definitely is confusing otherwise.}
% \ts{Right!}
The negative value of the N–N 2NN SRO parameter in Figure~\ref{fig2}(c) reflects the formation of 2NN N–N pairs, consistent with prior experimental and theoretical findings~\cite{gavrilyuk1990distribution}.

% The optimal CE model was employed to calculate the Warren-Cowley SRO parameters between N-metal and N-N pairs in the alloy system.
% To represent typical austenitic stainless steels, we selected an alloy composition of Fe$_{70}$Ni$_{10}$Cr$_{20}$, incorporating either 1 at.\% or 10 at.\% N atoms (N/M ratio) .
% In the 1 at.\% N case, the temperature dependence of the N–metal and N–N SRO parameters is shown in Figure~\ref{fig2}.
% The N-Cr 1NN SRO parameter in Figure~\ref{fig2}(a) exhibits a pronounced transition from negative to positive around 800 K.
% This behavior suggests a strong preference of N-Cr 1NN pairs at high temperatures, while Fe-N 1NN pairs are more favored at temperatures below 800 K.
% Conversely, in Figure~\ref{fig2}(b), the N–Cr 1NN SRO parameter remains negative below 1000 K, indicating order–disorder transition behavior.
% The negative value of the N–N 2NN SRO parameter in Figure~\ref{fig2}(c) reflects the formation of N–N 2NN pairs, consistent with both experimental and theoretical reports~\cite{gavrilyuk1990distribution}.

 %*

Monte Carlo snapshots at 1500 K, 1000 K, and 500 K are shown in Figures~\ref{fig2}(d,e,f), respectively. 
At 1500 K, N–Cr 1NN pairs outnumber N–Fe pairs despite Fe being significantly more abundant, highlighting the strong N–Cr affinity, consistent with the negative effective cluster interaction (ECI) for N–Cr 1NN pairs. As temperature decreases to 1000 K, N–Fe 1NN pairs become more prevalent. 
At 500 K, the local environment is dominated by N–Fe 1NN pairs, with only a few remaining N–Cr 1NN pairs. 
This trend is attributed to the emergence of Fe–Cr chemical ordering that favors Fe$_3$Cr-like structures (Figure~\ref{fig2}(f)), which suppress N–Cr 1NN pairing and instead promote N–Cr 2NN configurations. Additionally, the formation of dumbbell-like N–N 2NN structures is observed, consistent with the negative 2NN SRO parameter.

% Figure~\ref{fig2}(d),(e), and (f) present the MC snapshots at 1500 K, 1000 K, and 500 K, respectively.
% At 1500 K, N–Cr pairs are more abundant than N–Fe pairs, despite the significantly higher Fe concentration relative to Cr.
% This highlights the strong affinity between N and Cr atoms, which is again consistent with the negative ECI for N-Cr 1NN pairs.
% As the temperature decreases to 1000 K, the number of N–Fe 1NN pairs increases noticeably.
% At 500 K, N–Fe 1NN pairs dominate the local environment, while fewer N–Cr pairs remain.
% This trend suggests that at lower temperatures, the dominant Fe–Cr chemical interactions drive the formation of Fe$_3$Cr-like ordered structures (in Figure~\ref{fig2}(f)), which suppress the occurrence of N-Cr 1NN pairs while promoting the formation of N-Cr 2NN pairs.
% Simultaneously, the formation of N–N dumbbell-like structures is observed, aligning with the negative N–N 2NN SRO parameter.

\begin{figure*}[!hbtp] %*
\centering
\includegraphics[width=6in]{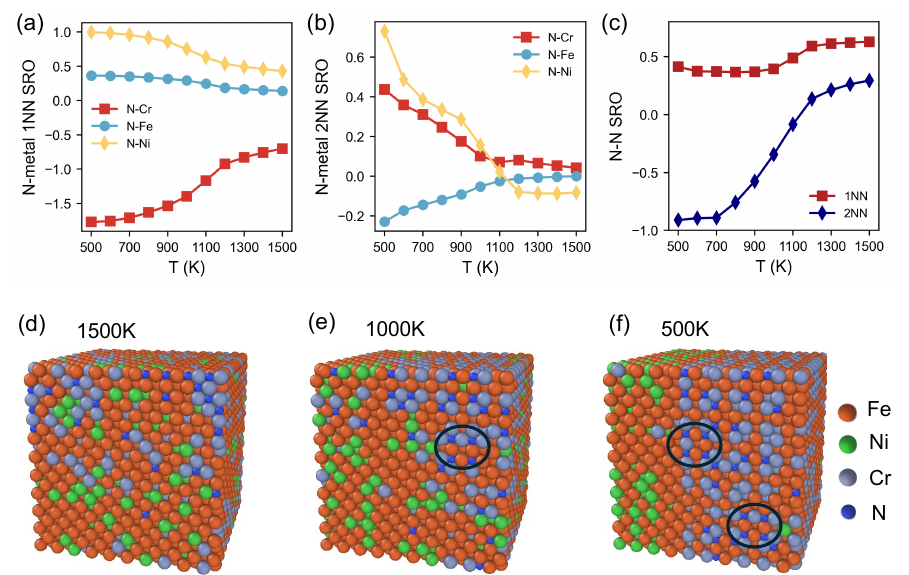}
\caption{ 
(a, b) First- and second-nearest-neighbor (1NN and 2NN) SRO parameters between nitrogen and metal atoms, and (c) 2NN N–N SRO parameter as a function of temperature for the 10 at.\% N case in Fe$_{70}$Ni$_{10}$Cr$_{20}$.
(d–f) Representative Monte Carlo snapshots at 1500 K, 1000 K, and 500 K, respectively, showing atomic distributions. Black circles in (e) and (f) highlight the formation of M$_4$N-like short-range ordered structures.
Fe, Ni, Cr, and N atoms are shown in red, green, blue, and dark blue, respectively.
}
\label{fig3}
\end{figure*} %*

The structural characteristics of high-N phases are particularly intriguing. 
When nitrogen concentrations approach the solubility limit (10 at.\%), Fe$_4$N-like structures and nanoscale Cr–N clusters have been experimentally observed in austenitic stainless steels\cite{xie2022nanosized,che2019co}.
In addition, nitrogen-driven local chemical order~\cite{he2023interstitial} and various nitrides~\cite{somers2013nitriding} have been reported in high-N alloys. 

To investigate this regime, we simulated Fe$_{70}$Ni$_{10}$Cr$_{20}$ with 10 at.\% nitrogen. 
The results are summarized in Figure~\ref{fig3}. 
Compared to the low-N case, the SRO trends become more pronounced. The N–Cr 1NN SRO parameter remains negative across all temperatures, while the 2NN parameter becomes positive, signifying a shift toward more extended N–Cr interactions. 
This behavior correlates with the development of M$_4$N-like SRO domains, as shown in the MC snapshots in Figures~\ref{fig3}(d) and (e). At 1500 K, localized Cr–N-rich regions already exist, and by 1000 K, the M$_4$N-like domains become well defined. 
At the same time, the N–N 2NN SRO parameter becomes increasingly negative, indicating the formation of N–N dumbbells mediated by the surrounding M$_4$N-like structure. 
By 500 K, these SRO domains evolve into long-range ordered (LRO) regions.

The formation of M$_4$N-like phases has been experimentally confirmed~\cite{che2019co}, and recent theoretical studies have shed light on the dynamics of LRO formation~\cite{che2022co,che2023co}. 
Nevertheless, the precise atomic arrangement of Fe, Ni, Cr, and N in these high-N phases remains elusive. 
The CE–MC simulations presented here, rooted in first-principles calculations, offer direct atomic-scale insights into the structure and evolution of both SRO and LRO. 
%\sout{Notably, the ordered regions exhibit a lamellar morphology, reminiscent of structures reported in high-entropy alloys~\cite{he2023interstitial}.} \jak{see my comment about about this reference.  When you say that you have a "lathe-like" structure in your observations, you're referring to the segregation of the Fe-Ni vs. the Fe-Cr-N, right? Is this a big enough unit cell to make that statement confidently?  Have you seen it consistently? Your following sentence is quite compelling, but at this lengthscale we'ver really shifted towards a precipitation mechanisms, right? and it's known, at least for Cr2N precipitates, that this is a discontinuous precipitation mechanism that often manifests in a lamellar structure, with acicular Cr2N precipitates in the gamma matrix. See this reference for some examples: https://doi.org/10.1080/14786430701805590 }
These results suggest that nitrogen not only stabilizes specific chemical environments but also promotes anisotropic ordering, potentially impacting microstructural and mechanical properties.

\subsection{The effect of N and Cr on SRO}

\begin{figure*}[!hbtp] %*
\centering
\includegraphics[width=3in]{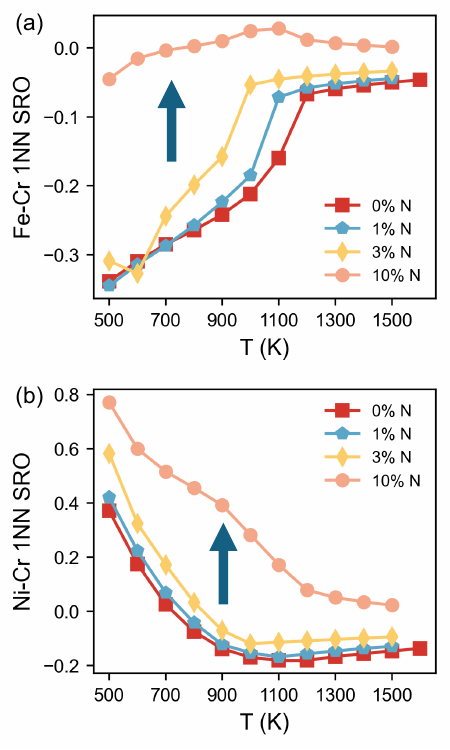}
\caption{ 
Effect of nitrogen concentration on 1NN (a) Fe–Cr and (b) Ni–Cr SRO parameters in the Fe$_{70}$Ni$_{10}$Cr$_{20}$ alloy. Nitrogen content is varied from 0 to 10 at.\%. Arrows indicate the significant shift in ordering behavior observed when the N concentration reaches 10 at.\%, reflecting the emergence of N–Cr dominated short-range order.
}
\label{fig4}
\end{figure*} %*

The influence of nitrogen incorporation on the intrinsic SRO in Fe$_{70}$Ni$_{10}$Cr$_{20}$ alloys is shown in Figure~\ref{fig4}, where the N concentration is varied across 0, 1, 3, and 10 at.\%. 
We focus here on the first-nearest-neighbor (1NN) SRO parameters for Fe–Cr and Ni–Cr pairs, while additional, less impactful SRO parameters are provided in ESI Figure S6.
At low N concentrations, the intrinsic SRO of austenitic stainless steels is only mildly perturbed. 
As the N content increases from 0 at.\% to 3 at.\%, the order–disorder transition temperature of Fe–Cr 1NN pairs decreases by approximately 200 K. 
However, the overall ordering tendencies are preserved: Fe–Cr 1NN pairs remain energetically favorable across the entire temperature range, while Ni–Cr 1NN pairs become increasingly unfavorable below $\sim$700 K.
In contrast, at 10 at.\% N, the SRO parameters exhibit marked changes due to the emergence of SRO domains. 
In this high-N regime, the Fe–Cr 1NN SRO approaches zero, and the Ni–Cr 1NN SRO remains positive across all temperatures. 
This behavior arises from the tendency of Cr atoms to segregate into N–Cr rich regions, reducing the likelihood of Fe–Cr or Ni–Cr pairing.

% The effect of N incorporation on the intrinsic SRO parameters of the Fe$_{70}$Ni$_{10}$Cr$_{20}$ alloy is presented in Figure~\ref{fig4}.
% The N concentration varies from 0, 1, 3, and 10 at.\%.
% Here, we focus on the Fe–Cr and Ni–Cr 1NN SRO parameters, with additional SRO parameters that are less significant provided in SI Figure S6.
% At low N concentrations, the intrinsic SRO of austenitic stainless steels is only mildly affected by N incorporation.
% As N concentration increase from 0 at.\% to 3 at.\%, the order-disorder transition temperature of Fe-Cr 1NN pairs decreases by around 200 K.
% However, the overall ordering tendencies remain largely unchanged: Fe–Cr 1NN pairs continue to be favored across the temperature range, while Ni–Cr 1NN pairs become increasingly unfavorable below 700 K.
% When the N concentration reaches 10 at.\%, the SRO parameters undergo significant changes due to the formation of SRO domains.
% Specifically, the Fe–Cr 1NN SRO approaches zero, and the Ni–Cr 1NN SRO remains positive across the entire temperature range.
% In this high N case, most Cr atoms are segregated within local N-Cr rich SRO regions, thus reducing the possibility of Fe-Cr and Ni-Cr pairs.

\begin{figure*}[!hbtp] %*
\centering
\includegraphics[width=6in]{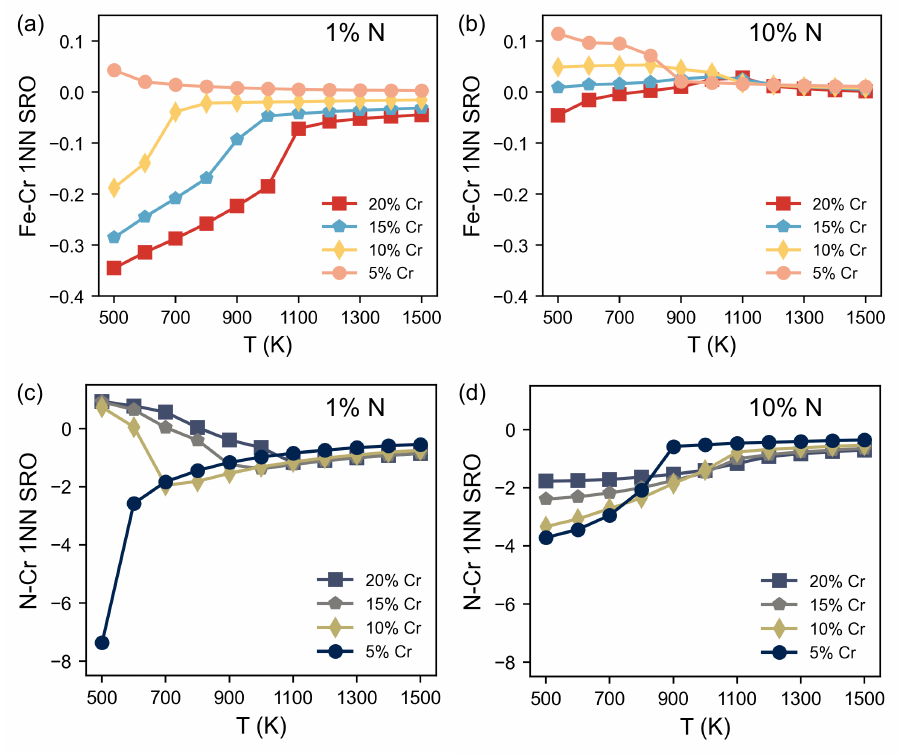}
\caption{ 
Influence of Cr concentration on 1NN SRO parameters at two different nitrogen contents. (a,b) Fe–Cr 1NN SRO for 1 at.\% and 10 at.\% N, respectively. (c,d) N–Cr 1NN SRO for 1 at.\% and 10 at.\% N, respectively. The Ni concentration is fixed at 10 at.\%, while Cr concentration varies from 5 at.\% to 20 at.\%. The emergence of strong N–Cr ordering at higher N content highlights the dominant role of N–Cr interactions in controlling short-range order under N-rich conditions.
}
\label{fig5}
\end{figure*} %*

To further assess the role of Cr content, we evaluated alloys with Cr concentrations ranging from 5 to 20 at.\%, keeping Ni fixed at 10 at.\%. 
Figure~\ref{fig5} presents the resulting Fe–Cr and N–Cr 1NN SRO parameters for both 1 at.\% and 10 at.\% N concentrations. 
As shown in Figure~\ref{fig5}(a), for the 1 at.\% N case, increasing the Cr content from 5 to 20 at.\% leads to a more negative Fe–Cr 1NN SRO, reflecting a stronger tendency for Fe–Cr pair formation. 
Concurrently, the N–Cr 1NN SRO (Figure~\ref{fig5}(c)) becomes less negative, indicating a reduced probability of N–Cr pair formation. 
This trend arises from the dominant influence of Fe–Cr chemical ordering at low N concentrations, consistent with observations in the previous section.

At 10 at.\% N, however, a different behavior emerges. 
As shown in Figure~\ref{fig5}(b), the Fe–Cr 1NN SRO is significantly suppressed relative to the 1 at.\% N case, while the N–Cr 1NN SRO (Figure~\ref{fig5}(d)) becomes strongly negative, indicating the formation of N–Cr enriched SRO domains. 
A comparison between Figures~\ref{fig5}(a, c) and \ref{fig5}(b, d) illustrates that, at high N concentrations, the sensitivity of SRO to alloy composition is diminished. 
In this regime, the strong affinity between N and Cr becomes the dominant driver of chemical ordering.

% To further assess the role of Cr content, we evaluated alloys with Cr concentrations ranging from 5 to 20 at.\%, keeping Ni fixed at 10 at.\%. 
% Figure~\ref{fig5} presents the resulting Fe–Cr and N–Cr 1NN SRO parameters for both 1 at.\% and 10 at.\% N concentrations. 
% As shown in Figure~\ref{fig5}(a), for the 1 at.\% N case, increasing the Cr content from 5 to 20 at.\% leads to a more negative Fe–Cr 1NN SRO, reflecting a stronger tendency for Fe–Cr pair formation. 
% Concurrently, the N–Cr 1NN SRO (Figure~\ref{fig5}(c)) becomes less negative, indicating a reduced probability of N–Cr pair formation. 
% This trend underscores the dominant influence of Fe–Cr chemical ordering at low N concentrations, consistent with observations in the previous section.
% At 10 at.\% N, however, a different behavior emerges. 
% As shown in Figure~\ref{fig5}(b), the Fe–Cr 1NN SRO is significantly suppressed relative to the 1 at.\% N case, while the N–Cr 1NN SRO (Figure~\ref{fig5}(d)) becomes strongly negative, indicating the formation of N–Cr enriched SRO domains. 
% A comparison between Figures~\ref{fig5}(a, c) and \ref{fig5}(b, d) illustrates that, at high N concentrations, the sensitivity of SRO to alloy composition is diminished. 
% In this regime, the strong affinity between N and Cr becomes the dominant driver of chemical ordering.

\section{Discussion}

\subsection{Impact of Magnetism on N-Related Short Range Order}

\begin{figure*}[!hbtp] %*
\centering
\includegraphics[width=6in]{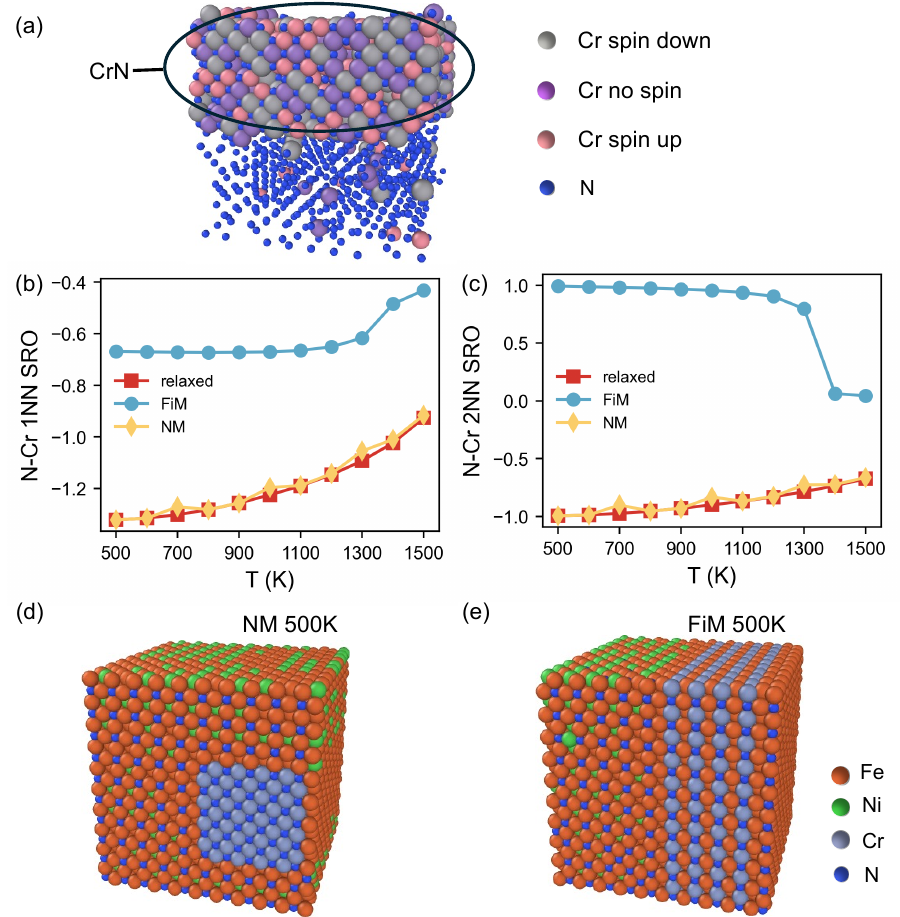}
\caption{ 
(a) A representative MC snapshot at 1500 K for the CrN nitride. Only Cr and N atoms are shown, while Fe and Ni atoms are hidden. The magnetic states of the Cr atoms are labeled with grey, purple, and pink for spin-down, no-spin, and spin-up, respectively. The number of Cr atoms that are spin up and spin down is almost equal. (b,c) The temperature dependence of N-Cr 1NN and 2NN SRO parameters for fully relaxed, nonmagnetic (NM), and pseudo-ferrimagnetic (FiM) states.  
(d,e) Representative MC snapshots at 500K from the NM and FiM simulations. Fe, Ni, Cr, and N atoms are marked with red, green, blue, and dark blue, respectively.
}
\label{fig6}
\end{figure*}

Magnetic disorder has been reported to contribute to the thermodynamic stability of cubic B1-type CrN in DFT simulations \cite{alling2010effect}.
In this study, we incorporate magnetic degrees of freedom into a spin CE model to investigate how magnetism influences SRO in high-N Fe-based alloys.
Here, the constrained spin CE was employed because it closely reproduces key magnetic benchmarks and is expected to provide a more reliable description of magnetic effects on SRO.
We examine the Fe$_{70}$Ni$_{10}$Cr$_{20}$ alloy with 40 at.\% N, a concentration at which nitride formation occurs in the present Monte Carlo simulations, specifically to study the effect of magnetism on CrN. 
%\sout{Specifically, we focus on the Fe$_{70}$Ni$_{10}$Cr$_{20}$ alloy with 40 at.\% N, a concentration where CrN formation has been experimentally observed~\cite{somers2013nitriding}.} \jak{Can you confirm that refs 44 and 45 are the best citations here? 44 mentions CrN, but not extensively and I'm having trouble finding a version of ref 45 that I can access? I think that these references are important, as CrN precipitation below 1000C is sometimes observed, but is perhaps disputed? I don't disagree with your results, but they need some better context, eg. why is it relevant at 1500k?}
A representative Monte Carlo snapshot at 1500 K, in which both chemical and magnetic configurations are allowed to relax, illustrates the formation of CrN-rich domains (see Figure~\ref{fig6}(a)).
Within these domains, Cr spins appear randomly oriented, resembling a paramagnetic or nonmagnetic state.
This suggests that magnetic disorder may facilitate CrN ordering.

 %*

% Magnetic disorder has been reported to contribute to the thermodynamic stability of the cubic B1-type CrN nitride~\cite{alling2010effect}.
% In this study, magnetic degrees of freedom are incorporated into the spin CE model to investigate the effect of magnetism on SRO in high-N alloys.
% Specifically, we focus on the Fe${70}$Ni${10}$Cr$_{20}$ alloy with a nitrogen concentration of 40 at.\%, a level at which nitride formation has been observed experimentally~\cite{somers2013nitriding,somers2024surface}.
% A representative snapshot from the MC simulation at 1500 K, allowing full relaxation of both chemical and magnetic configurations, reveals the formation of CrN domains(see SI Figure S8(a)).
% The spins of Cr atoms within the nitride region appear randomly distributed, resembling a paramagnetic or nonmagnetic state. 

To assess this hypothesis, we performed additional MC simulations under two constrained magnetism scenarios: (1) a nonmagnetic (NM) state where all metal atoms have zero spin, and (2) a pseudo-ferrimagnetic (FiM) state where Fe and Ni spins are fixed as up, and Cr spins as down. 
%\ee{Here, the constrained spin CE was used because .... }
The resulting N–Cr SRO parameters are compared in Figures~\ref{fig6}(b,c).
The NM state produces similar N–Cr SRO behavior as the fully relaxed case, supporting the idea that magnetic disorder promotes CrN formation.
In contrast, the FiM state leads to significantly weaker N–Cr ordering, as reflected in the reduced magnitude of the SRO parameters.
Snapshots of the atomic configurations (Figures~\ref{fig6}(d,e)) further show that the FiM state favors Fe–Cr ordering and suppresses the formation of B1-type CrN-like domains. 
Although austenitic stainless steels are typically nonmagnetic or paramagnetic at room temperature, these results indicate that magnetic interactions—especially competing spin orientations—can enhance or hinder CrN formation in nitrogen-enriched alloys.

% To assess whether this magnetic disorder is a driving force for nitride formation, additional MC simulations with constrained magnetic configurations were performed.
% Two scenarios were considered: a nonmagnetic (NM) state where all metal atoms have zero spin, and a pseudo-ferrimagnetic (FiM) state where Fe and Ni atoms are constrained to spin-up and Cr atoms to spin-down.
% The resulting N–Cr SRO parameters under these magnetic constraints are compared to the fully relaxed case shown in SI Figures S8(b,c).
% The NM state exhibits similar N–Cr ordering behaviors to the fully relaxed case, suggesting that magnetic disorder facilitates nitride formation.
% In contrast, the FiM state reduces the tendency for N–Cr ordering, as indicated by the smaller magnitude of the SRO parameters.
% The MC snapshots in SI Figures S8(d,e) clearly show that the FiM state promotes Fe–Cr ordering and suppresses the development of B1-type CrN nitride compared to the NM case. 
% Although austenitic stainless steel is typically nonmagnetic or paramagnetic at room temperature, these results suggest that alternating magnetic interactions can enhance or impede CrN formation in high-N alloys.

\subsection{Kinetic Constraints on N-Related Short-Range Order}

The previous analysis assumes that the system reaches thermodynamic equilibrium at each temperature. 
However, this assumption becomes less valid at low temperatures, where the diffusion of metal atoms is significantly limited. 
To assess how sluggish kinetics influences SRO  development, we conducted MC simulations in which the metal sublattice was held fixed, allowing only nitrogen atoms to exchange positions.

The results, summarized in SI Figure S7, highlight the strong influence of the initial metal configuration. 
Specifically, we compared two scenarios: one with an ordered metal lattice taken from MC simulations at 500 K, and another with a disordered configuration from 1500 K. 
In the disordered case, Cr–N SRO still emerges, indicating that N–Cr interactions are strong enough to drive local ordering even in the absence of metal atom rearrangement. 
In contrast, the pre-ordered metal lattice favors the formation of N–Fe pairs, as the existing Fe–Cr ordering restricts N access to Cr neighbors. 
%\jak{this is a nice result!}

This comparison highlights that pre-existing SRO among metal atoms can strongly modulate the development of N-related SRO. 
In practice, nitriding treatments are often performed near 1000 K, a temperature generally sufficient to enable significant metal diffusion. 
Nonetheless, under rapid cooling or short annealing times, non-equilibrium effects may persist, leading to SRO states that reflect a combination of thermodynamic preferences and kinetic constraints.  
%\jak{also a really key observation!}

% The above analysis is based on the assumption of thermodynamic equilibrium at a given temperature.
% However, at low temperatures, the diffusion of metal atoms is significantly reduced.
% To illustrate the impact of sluggish kinetics on the development of SRO, we performed MC simulations where the metal sublattice was fixed, allowing only N atoms to swap positions.
% The influence of the initial alloy configuration is evident, as shown in SI Figure S7.
% The ordered and random alloy configurations are taken from MC simulations of the pure Fe$_{70}$Ni$_{10}$Cr$_{20}$ alloy at 500 K and 1500 K, respectively.
% When the metal lattice is fixed in a random configuration, Cr–N SRO can still develop.
% In contrast, if the underlying metal structure is already ordered, N–Fe pairs are more favored.
% This comparison highlights that the pre-existing SRO in the metal lattice strongly affects the formation of N-related SRO.
% % In a random alloy, even with limited diffusion, N–Cr SRO may still emerge 
% % However, certain local chemical environments can impede the formation of N–Cr pairs.
% In reality, the nitriding process is usually conducted at around 1000 K, which should be sufficient for metal atom diffusion.

\subsection{Advantages and Shortcomings of the Current Model}

Magnetism is known to play a significant role in chemical SRO across various alloy systems, including NiCrCo medium-entropy alloys and CrFeCoNi high-entropy alloys~\cite{niu2015spin,walsh2021magnetically,woodgate2023interplay}.
In particular, the magnetic interactions involving Cr atoms strongly influence their ordering behavior.
For austenitic stainless steels, finite-temperature magnetism has been shown to markedly affect Fe–Cr SRO~\cite{ruban2016atomic,su2024first}, underscoring the importance of accounting for magnetism when modeling chemical order.
In this study, we employ a spin CE model that explicitly incorporates magnetic degrees of freedom, following our previous methodology.
%Our Monte Carlo simulations allow both spin flips and atom swaps, enabling a realistic description of the coupling between magnetism and chemical ordering in high-N austenitic stainless steels.
Compared to prior MC studies that neglect magnetism, the approach offers improved fidelity in capturing the magnetic contributions to SRO.
%It also provides a significant computational advantage over density functional theory (DFT)–based methods:
The use of statistical sampling over large supercells allows us to efficiently explore temperature and compositional effects, rather than relying on isolated DFT configurations~\cite{tong2019short}.

However, the current model has limitations.
Strain effects are not explicitly treated; the spin CE model is constructed from fully relaxed DFT structures, meaning that elastic degrees of freedom are only implicitly encoded.
By contrast, mixed-space cluster expansion (MSCE) approaches~\cite{holliger2011reciprocal,wang2023generalization} explicitly include strain by working in reciprocal space.
As a result, our results correspond to idealized, stress-free conditions.
Future work could extend this framework to explore strain-driven effects in N-rich austenitic alloys—particularly at extreme N concentrations where lattice distortion becomes significant. 
Moreover, the model system here is constrained to the FCC lattice, which limits access to martensite or disordered structures. 
Previous experiments have characterized the structure of Cr$_2$N precipitates in high-N alloys~\cite{lee2005crystal} and documented the evolution of the hexagonal Cr$_2$N structure~\cite{wan2015nitrogen}. However, these phases are not readily accessible within the current model. 
Additional investigations are warranted to assess vibrational entropy and phase stability across different structural motifs, such as martensite and Cr-based nitrides.

\section{Conclusions}

We present a spin cluster expansion and Monte Carlo approach to investigate the effects of composition, magnetism, and temperature on both intrinsic and N-related SRO in high-N austenitic stainless steels.
The results reveal that nitrogen concentration and alloy composition significantly modulate chemical ordering.
At low N concentrations, N–Cr 1NN pairs are favored at high temperatures, while N–N 2NN correlations emerge at intermediate temperatures.
At high N levels, M$_4$N-like SRO and even long-range ordered domains form, particularly in Cr-rich regions.
The influence of Cr content is pronounced when nitrogen is dilute but diminishes near the solubility limit, where strong N–Cr interactions dominate.
We also find that sluggish metal diffusion at low temperatures can kinetically trap different N-related SRO configurations depending on the underlying metal lattice.
Magnetism further modulates chemical ordering: disordered spin states facilitate CrN domain formation, while ferrimagnetic alignment suppresses it. 
This framework provides a computationally efficient route to predict N-driven ordering in stainless steels and offers mechanistic insights relevant to nitriding process design.
Controlling factors such as nitriding temperature, nitrogen concentration profiles, and local composition will be key to tailoring microstructures for improved performance.

% In conclusion, we present a CE-MC approach that enables the investigation of composition, magnetism, and temperature effects on both N-related and intrinsic SRO in high-N austenitic stainless steels.
% Our findings demonstrate that alloy composition and N concentration significantly influence the formation of N-related SRO structures.
% At low N concentrations, N–Cr 1NN pairs are favored at high temperatures, while N–N 2NN pairs emerge at intermediate temperatures.
% At higher N concentrations, M$_4$N-like SRO or LRO structures are observed.
% Cr content strongly affects SRO when N is dilute, but its influence diminishes as N approaches the solubility limit.
% Kinetic effects, such as sluggish metal diffusion at low temperatures, alter N distribution and related SRO. 
% Magnetism also plays a crucial role: disordered magnetic states can promote nitride formation, which is captured in our simulations.
% This study offers an efficient way to predict the N-driven ordering behavior in high-N austenitic stainless steels, and provides insights into optimizing the nitriding process.
% Key factors such as nitriding temperature, nitrogen concentration profile, and local chemical environment all play crucial roles in determining the microstructural evolution.

\section{Acknowledgments}
The authors acknowledge support from the US Department of Energy H2@Scale program, through award DE-EE0008832. 
This work was partially supported by the US Department of Energy/National Nuclear Security Administration through the Chicago/DOE Alliance Center, cooperative agreements DE-NA0003975 and DE-NA0004153.  %\ee{[See if Jessica has any other funding to acknowledge?]}\jak{[acknowledging the H2@scale funding is just fine for me, thanks for checking!]}
This work used PSC Bridges-2 at the Pittsburgh Supercomputing Center through allocation MAT220011 from the Advanced Cyberinfrastructure Coordination Ecosystem: Services \& Support (ACCESS) program, which is supported by National Science Foundation grants \#2138259, \#2138286, \#2138307, \#2137603, and \#2138296.

% Reviewer:
% Experimental:
% Valentin Gavriljuk; Marcel A.J. Somers;
% Theoretical: Axel VdW...

\newpage
\bibliography{mybib.bib}

\end{document}